# EXTENDED WIGNER FUNCTION FOR THE HARMONIC OSCILLATOR IN THE PHASE SPACE

**E.E. Perepelkin[a], B.I. Sadovnikov[a], N.G. Inozemtseva[b], E.V. Burlakov[a]**

[a] *Faculty of Physics, Lomonosov Moscow State University, Moscow, 119991 Russia*
E. Perepelkin e-mail: pevgeny@jinr.ru, B. Sadovnikov e-mail: sadovnikov@phys.msu.ru
[b] *Dubna State University, Moscow region, Moscow,141980 Russia*
e-mail: nginozv@mail.ru

**Abstract**

New time dependent Wigner functions for the quantum harmonic oscillator have been obtained in this work. The Moyal equation for the harmonic oscillator has been presented as the wave equation of a 2D membrane in the phase plane. The values of the Wigner function are equal to the deviation values of the points on the surface of the membrane from the equilibrium state. The positive and negative values of the Wigner function correspond to the direction of the deviation from the equilibrium state. As an example, a time dependent Wigner function corresponding to the standing wave of quasi-probability density arising in the phase plane is considered.

**Introduction**

In 1932, in the works of E. Wigner and H. Weyl [1,2], a quasi-probability function $W(x,p,t)$ was introduced to describe a quantum system in the phase space. The name «quasi-probability function» is due to the following features. First, considering a quantum system in the phase space with the Heisenberg uncertainty principle is a sophisticated problem. Secondly, the Wigner function $W$ can take negative values [3-6], which looks strange from the standpoint of the classical apparatus of probability theory. Despite the apparent inconsistency of the introduction of the Wigner function, many works have been devoted to the development of the mathematical apparatus of quantum mechanics in the phase space [7-21]. The Wigner function has been widely used in quantum tomography [22-25], quantum communication and cryptography [26-27], quantum informatics [28-30] and in signal processing tasks [31-33].

The evolution of the Wigner function $W(x,p,t)$ is determined by the Moyal equation [34-36], which is an extended analogue of the Liouville equation. The Moyal equation differs from the Liouville equation by a nonzero right-hand side, which is associated with dissipations in quantum systems [37-40]. A quantum harmonic oscillator is the simplest system for which an exact expression for the Wigner function $W_n(x,p)$ is obtained, where $n$ is the state number [41]. When considering the harmonic oscillator, the right-hand side of the Moyal equation is equal to zero, and the Moyal equation transforms into the Liouville equation. Thus, the function $W_n(x,p)$ is a stationary solution of the Moyal equation for the harmonic oscillator.

In this work, new ***time dependent*** Wigner functions $W^{(n)}(x,p,t)$ for the harmonic oscillator have been found. It turned out that the Moyal equation can be written as the wave equation of a 2D membrane. A phase plane has been used as a membrane. The values of the Wigner function correspond to the deviations of the membrane points from the equilibrium state. The positive and negative values of the quasi-probability density can be interpreted as deviations from the equilibrium position in one or the other direction.

The work has the following structure. In §1, we obtain the new class of time dependent Wigner functions $W^{(n)}(x,p,t)$ for the harmonic oscillator. The main theorems are formulated. The resulting expressions for $W^{(n)}(x,p,t)$ can have non-smooth points at the origin of the phase plane (a state with zero energy $\varepsilon = 0$, where $\varepsilon = \frac{1}{\hbar\omega}\left(\frac{p^2}{2m} + \frac{m\omega^2 x^2}{2}\right)$). The presence of a non-smooth point of the function $W^{(n)}(x,p,t)$ at the origin introduces uncertainty into the probability value of a state with zero energy. In the particular case, the functions $W^{(n)}(x,p,t)$ transform into the well-known Wigner functions $W_n(x,p)$. When integrating over the coordinate space and momenta, new functions $W^{(n)}(x,p,t)$ give the corresponding known probability densities of coordinates $|\Psi_n(x)|^2$ and momenta $|\tilde{\Psi}_n(p)|^2$ of the harmonic oscillator. The energy spectrum corresponds to a harmonic oscillator $\langle\langle\varepsilon\rangle\rangle_n = \hbar\omega(n+1/2)$, $n \in \mathbb{N}$, where averaging is performed with the density of quasi-probabilities $W^{(n)}(x,p,t)$ over the entire phase space.

In §2, we consider an example of a time dependent Wigner function $W_\ell^{(n)}(x,p,t)$ for the harmonic oscillator. The considered example corresponds to a standing wave, which arises along the circular phase trajectories of the oscillator. The phase velocity of the wave is equal to the frequency $\omega$ of the oscillator $\omega$ (angular velocity). The frequency of the wave $\Omega_\ell$ is a multiple of the frequency $\omega$, that is $\Omega_\ell = 2\omega\ell$, where $\ell \in \mathbb{N}$. The properties of the obtained solutions are considered. The ground state ($n=0$) in this case has a positive function $W_\ell^{(0)}(x,p,t)$ different from the Gaussian distribution, which is consistent with the Hudson's theorem [42], due to the presence of a non-smooth point at the origin.

In conclusion, the main results of the work are considered. The Appendix provides the proofs of the theorems.

**§1 Wave equation**

Let us consider the modified Vlasov equation [43, 44] for the distribution density function $f(x,v,t)$

$$\frac{\partial f}{\partial t} + v\frac{\partial f}{\partial x} + \frac{\partial}{\partial v}\left(\langle \dot{v}\rangle f\right) = 0. \tag{1.1}$$

Taking into consideration the Vlasov-Moyal approximation [39],

$$\langle \dot{v}\rangle = \sum_{n=0}^{+\infty} \frac{(-1)^{n+1}(\hbar/2)^{2n}}{m^{2n+1}(2n+1)!} \frac{\partial^{2n+1}U}{\partial x^{2n+1}} \frac{1}{f}\frac{\partial^{2n} f}{\partial v^{2n}}, \tag{1.2}$$

the equation (1.1) becomes the Moyal equation for the Wigner function $W$:

$$\frac{\partial W}{\partial t} + \frac{p}{m}\frac{\partial W}{\partial x} - \frac{\partial U}{\partial x}\frac{\partial W}{\partial p} = \sum_{n=1}^{+\infty} \frac{(-1)^n(\hbar/2)^{2n}}{(2n+1)!}\frac{\partial^{2n+1}U}{\partial x^{2n+1}}\frac{\partial^{2n+1}W}{\partial p^{2n+1}}, \tag{1.3}$$

where $f(x,v,t) = W(x,p,t)$; $U(x)$ – is the potential from the Schrödinger equation. For a harmonic oscillator with a potential

$$U_\alpha(x) = \frac{m\omega^2 x^2}{2} + \alpha x + \frac{\alpha^2}{2m\omega^2} = \frac{m\omega^2 \bar{x}^2}{2}, \quad \bar{x} = x + \frac{\alpha}{m\omega^2},$$

the Moyal equation (1.3) and the Liouville equation coincide and are of the form:

$$\frac{\partial W}{\partial t} + \frac{p}{m}\frac{\partial W}{\partial x} - (m\omega^2 x + \alpha)\frac{\partial W}{\partial p} = 0, \quad \frac{\partial f}{\partial t} + v\frac{\partial f}{\partial x} - \left(\omega^2 x + \frac{\alpha}{m}\right)\frac{\partial f}{\partial v} = 0. \qquad (1.4)$$

***Theorem 1***. *The equation (1.4) can be presented as a wave equation*

$$W_{tt} = \omega^2 W_{\varphi\varphi}, \qquad (1.5)$$

*where*

$$W(\rho,\varphi,t) = W\left(\frac{\rho}{\omega}\cos\varphi - \frac{\alpha}{m\omega^2}, m\rho\sin\varphi, t\right).$$

The equation (1.5) is a wave equation in which the function W corresponds to the value of the deviation relative to the equilibrium state. The deviation W from the equilibrium state can take positive and negative values. Therefore, the positive and negative values of the Wigner function can be interpreted as positive and negative deviations from the equilibrium state during the propagation of the probability density wave.

The value $\omega$ in the equation (1.5) corresponds to the phase velocity of the wave, which propagates around the circumference in the phase plane $(\bar{x}, p)$. The solution of the equation (1.5) can be represented in the form

$$W(\rho,\varphi,t) = F(\rho,\xi(\varphi,t)) + G(\rho,\eta(\varphi,t)), \qquad (1.6)$$
$$\xi(\varphi,t) = \Omega t - \kappa\varphi, \quad \eta(\varphi,t) = \Omega t + \kappa\varphi, \quad \Omega = \omega\kappa,$$

where $F,G$ are certain arbitrary functions; $\xi(\varphi,t)$ and $\eta(\varphi,t)$ are characteristics along which the solution W has a constant value; the value $\kappa$ corresponds to the wave number; the signs «±» correspond to the clockwise and counter-clockwise direction of the wave around the origin in the phase plane $(\bar{x}, p)$.

**Remark**

As the angle $\varphi = \arctg\frac{v}{u} = \arctg\frac{p}{m\omega\bar{x}}$ corresponds to the angle in the phase plane $(\bar{x}, p)$, the solution (1.6) should satisfy the following periodic condition at every instant:

$$W(\rho,0,t) = W(\rho,2\pi,t) \Rightarrow F(\rho,\Omega t) + G(\rho,\Omega t) = F(\rho,\Omega t - 2\pi\kappa) + G(\rho,\Omega t + 2\pi\kappa), \quad (1.7)$$

i.e. the functions $F,G$ must be $2\pi\kappa$ periodic in the second argument. Note that according to (A.1), (A.3), the value $\rho$ corresponds to the energy of the harmonic oscillator:

$$\rho^2 = u^2 + v^2 = \omega^2 \bar{x}^2 + \frac{p^2}{m^2} = \frac{2}{m}\left(\frac{p^2}{2m} + \frac{m\omega^2 \bar{x}^2}{2}\right) = 2\frac{\hbar\omega}{m}\varepsilon(\bar{x}, p), \tag{1.8}$$

$$\varepsilon(\bar{x}, p) = \frac{1}{\hbar\omega}\left(\frac{p^2}{2m} + \frac{m\omega^2 \bar{x}^2}{2}\right) = \frac{1}{\hbar\omega}\left(\frac{p^2}{2m} + U_\alpha(x)\right).$$

On the other hand, the Wigner function for the quantum harmonic oscillator is known explicitly:

$$W_n(\bar{x}, p) = \frac{(-1)^n}{\pi\hbar} e^{-2\varepsilon(\bar{x},p)} L_n(4\varepsilon(\bar{x}, p)), \tag{1.9}$$

where $L_n$ are Laguerre polynomials. The solution (1.9) of the Moyal equation (1.4) is stationary and independent of time but depends only on the energy $\varepsilon(\bar{x}, p)$, i.e., according to (1.8), on the radius $\rho$. From a comparison of (1.7) and (1.9), it follows that the time dependent solution of the Moyal/Vlasov equation with the potential $U_\alpha(x)$ can be written in the factorized form:

$$W^{(n)}(\rho, \varphi, t) = N\frac{(-1)^n}{\pi\hbar} e^{-\frac{m}{\hbar\omega}\rho^2} L_n\left(\frac{2m}{\hbar\omega}\rho^2\right)\left[C + f(\Omega t + \kappa\varphi) + g(\Omega t - \kappa\varphi)\right], \tag{1.10}$$

where $f, g$ are certain $2\pi\kappa$ periodic functions and $C$ is certain constant value. The invariable $N$ is a normalization factor.

***Lemma 1.*** *In the expression (1.10), the normalization factor $N$ is of the form:*

$$N = \frac{1}{C + \langle f \rangle + \langle g \rangle}, \tag{1.11}$$

*where $\langle f \rangle$ and $\langle g \rangle$ are mean values of the functions $f, g$ along the interval $[0, 2\pi]$.*

**Remark**

Note that the normalization factor $N$ (1.11) will not depend on the time $t$. The value $\Omega t$ in (1.10) corresponds to the phase progression and gives a rotation around the origin $(\bar{x}, p)$ of the distributions $f, g$ that are $2\pi\kappa$ periodic.

The functions $f, g$ in the expression (1.10) are determined by the initial conditions $W_0^{(n)}(\rho, \varphi)$, for example:

$$W^{(n)}(\rho, \varphi, 0) = W_0^{(n)}(\rho, \varphi) = N\frac{(-1)^n}{\pi\hbar} e^{-\frac{m}{\hbar\omega}\rho^2} L_n\left(\frac{2m}{\hbar\omega}\rho^2\right)\left[C + f(\kappa\varphi) + g(-\kappa\varphi)\right]. \tag{1.12}$$

**Remark**

The simplest example of a solution of the form (1.10) is the case $f + g = 0$, $C = 1$ corresponding to the stationary solution (1.9). In the general case, the solution (1.10) is non-smooth at the origin of the phase plane $(\bar{x}, p)$ at $\rho = 0$, i.e., at zero energy. Thus, a state with zero energy is unstable, and at the slightest fluctuations, a sharp change in the density of the quasi-probabilities $W^{(n)}$ occurs.

When integrating the Wigner function $W_n(\bar{x}, p)$ over the momentum and coordinate, positive functions should be obtained:

$$\int_{-\infty}^{+\infty} W_n(\bar{x}, p)dp = |\Psi_n(\bar{x})|^2 = \frac{1}{2^n n!}\sqrt{\frac{m\omega}{\pi\hbar}}e^{-\frac{m\omega x^2}{\hbar}}H_n^2\left(\sqrt{\frac{m\omega}{\pi\hbar}}\bar{x}\right), \quad (1.13)$$

$$\int_{-\infty}^{+\infty} W_n(\bar{x}, p)dx = |\tilde{\Psi}_n(p)|^2,$$

where $\Psi_n(\bar{x})$ is a coordinate representation of the wave function, and $\tilde{\Psi}_n(p)$ is a momentum representation of the wave function of the quantum harmonic oscillator. The function $|\Psi_n(\bar{x})|^2$ corresponds to the probability density of the coordinate distributions, and the function $|\tilde{\Psi}_n(p)|^2$ corresponds to the probability density of the momentum distribution for the quantum harmonic oscillator.

Integrating the expression (1.10) over the coordinate and momentum, we obtain:

$$\int_{-\infty}^{+\infty} W^{(n)}(\bar{x}, p, t)d\bar{x} = CN|\tilde{\Psi}_n(p)|^2 + N\langle\Phi\rangle_{x,n}(p, t), \quad (1.14)$$

$$\int_{-\infty}^{+\infty} W^{(n)}(\bar{x}, p, t)dp = CN|\Psi_n(\bar{x})|^2 + N\langle\Phi\rangle_{p,n}(\bar{x}, t),$$

where

$$\langle\Phi\rangle_x(p, t) = \int_{-\infty}^{+\infty} \left[f(\Omega t + \kappa\varphi) + g(\Omega t - \kappa\varphi)\right]W_n(\bar{x}, p)d\bar{x}, \quad (1.15)$$

$$\langle\Phi\rangle_p(\bar{x}, t) = \int_{-\infty}^{+\infty} \left[f(\Omega t + \kappa\varphi) + g(\Omega t - \kappa\varphi)\right]W_n(\bar{x}, p)dp.$$

The functions $\langle\Phi\rangle_{x,n}$ and $\langle\Phi\rangle_{p,n}$ are the averaged functions for the function $\Phi(\bar{x}, p, t) = f(\eta) + g(\xi)$ over the coordinate and momentum, respectively. Averaging (1.15) of the function $\Phi$ is performed with the Wigner density of the quasi-probabilities $W_n(\bar{x}, p)$ (1.9). The functions $f, g$ are arbitrary periodic functions, therefore, the functions $\langle\Phi\rangle_{x,n}, \langle\Phi\rangle_{p,n}$ can take on different values. The consistency of the expressions (1.13) and (1.14) is determined by the following theorem.

**Theorem 2.** *Let the function $\Phi(\bar{x}, p, t)$ be odd in respect to the variables $\bar{x}$ and $p$, then the function $W^{(n)}(\bar{x}, p, t)$ (1.10) satisfies the conditions:*

$$\int_{-\infty}^{+\infty} W^{(n)}(\bar{x}, p, t)d\bar{x} = |\tilde{\Psi}_n(p)|^2, \qquad \int_{-\infty}^{+\infty} W^{(n)}(\bar{x}, p, t)dp = |\Psi_n(\bar{x})|^2, \quad (1.16)$$

The fulfillment of the conditions (1.16) means that the constructed function $W^{(n)}(\bar{x},p,t)$ (1.10) can be considered as a function of the quasi-probability density for the harmonic oscillator in the phase space.

***Definition 1****. Let the function $\Phi(\bar{x},p,t)$ be odd in respect to the variables $\bar{x}$ and $p$, then the corresponding function $W^{(n)}(\bar{x},p,t)$ (1.10) is called an extended Wigner function for the harmonic oscillator in the phase space.*

***Theorem 3****. The energy levels $\langle\langle\varepsilon_n\rangle\rangle$, corresponding to the extended Wigner function $W^{(n)}$ are of the form:*

$$\langle\langle\varepsilon_n\rangle\rangle = \hbar\omega \int_{-\infty}^{+\infty}\int_{-\infty}^{+\infty} \varepsilon(\bar{x},p) W^{(n)}(\bar{x},p) d\bar{x}dp = \hbar\omega\left(n+\frac{1}{2}\right). \quad (1.17)$$

**Remark**

The expression (1.17) coincides with the energy spectrum of the harmonic oscillator. Thus, the presence of the time dependent part $f(\eta)+g(\xi)$ in the expression (1.10) has no effects on the energy spectrum of the harmonic oscillator.

### §2 Example of an extended Wigner function

Let us consider model solutions of the Moyal equation (1.4) in the form of the extended Wigner function (1.10) for the quantum harmonic oscillator. In §1, it has been shown that the function $\Phi(\bar{x},p,t) = f(\eta)+g(\xi)$ must be odd in respect to the variables $\bar{x}$, $p$ and $2\pi\kappa$ periodic. As an example, we consider the representation of the function $\Phi$ as a superposition of two counterpropagating waves giving a standing wave:

$$\Phi(\bar{x},p,t) = g(\Omega t+\kappa\varphi) + f(\Omega t-\kappa\varphi) = A\sin(\Omega t+\kappa\varphi) - A\sin(\Omega t-\kappa\varphi),$$
$$\Phi(\bar{x},p,t) = \bar{\Phi}(\varphi,t) = 2A\cos(\Omega t)\sin(\kappa\varphi), \quad (2.1)$$

where $A$ is constant. The function (2.1) satisfies the periodicity condition $\bar{\Phi}(\varphi+2\pi\kappa,t) = \bar{\Phi}(\varphi,t)$. The function (2.1) is odd in respect to the variable $p$, since the substitution of $p$ for $-p$ corresponds to the substitution of $\varphi$ for $-\varphi$. The oddness of the function (2.1) in respect to the variable $\bar{x}$ is possible at $\kappa=2\ell$, $\ell\in\mathbb{N}$. The condition for the wave number to be $\kappa$ gives the restriction on the relation (1.6) of the frequencies $\Omega$ and $\omega$, i.e.

$$\Omega_\ell = 2\omega\ell. \quad (2.2)$$

The frequency $\omega$ corresponds to the frequency of the harmonic oscillator and determines the angular velocity, which the running waves have in the phase plane along the circumferences around the origin. From the expression (2.2) it follows that the frequency $\Omega$ of the standing wave (2.1) must be a multiple of the frequency $\omega$.

As a result, the conditions of Theorem 2 are satisfied for the function (2.1), and we can write the extended Wigner function $W^{(n)}(\bar{x},p,t)$ for the harmonic oscillator in the form (1.10):

$$W_\ell^{(n)}(\rho,\varphi,t) = \frac{(-1)^n}{\pi\hbar} e^{-\frac{m}{\hbar\omega}\rho^2} L_n\left(\frac{2m}{\hbar\omega}\rho^2\right)\left[1 + 2\frac{A}{C}\cos(2\omega\ell t)\sin(2\ell\varphi)\right], \qquad (2.3)$$

it is taken into account here that $N = 1/C$. For $A = 0$, the expression (2.3) transforms into a ordinary stationary Wigner function for the harmonic oscillator $W_n$ (1.9). From the expression (2.3) it follows that for each state $n$ of the harmonic oscillator there is an additional discrete set of functions determined by an index $\ell$. The nodes and antinodes of the standing wave will be located at the following angles in the phase plane:

$$\varphi_{n,\ell}^{(node)} = \frac{\pi n}{2\ell}, \quad \varphi_{n,\ell}^{(anti-node)} = \frac{\pi(2n+1)}{4\ell}. \qquad (2.4)$$

From the analysis of the expression (2.3) it follows that the Wigner function $W_n$ is a certain «average / equilibrium» function, in respect to which the oscillations of the function $W_\ell^{(n)}$ occur with a frequency $\Omega$ and period $T = 2\pi/\Omega$.

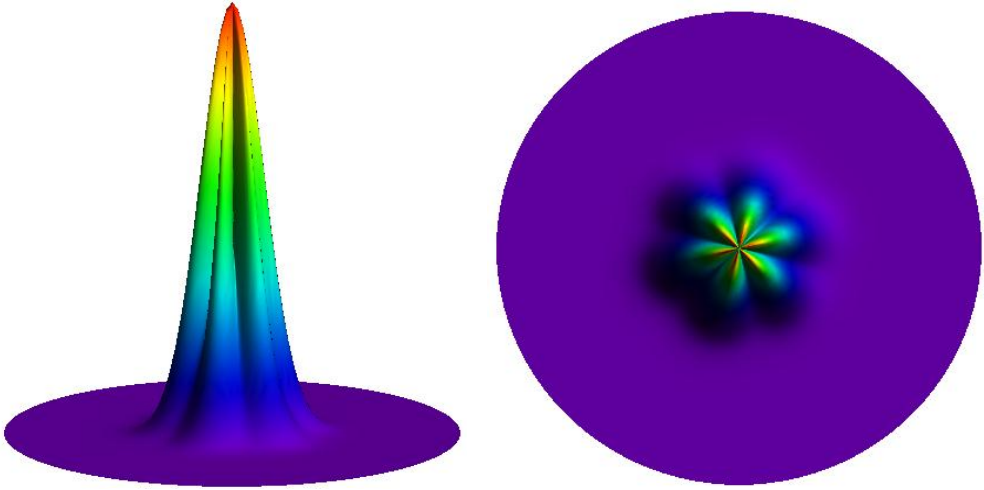

Figure 1. Graph of $W_3^{(0)}(\rho,\varphi,0) = W_3^{(0)}(\rho,\varphi,T)$

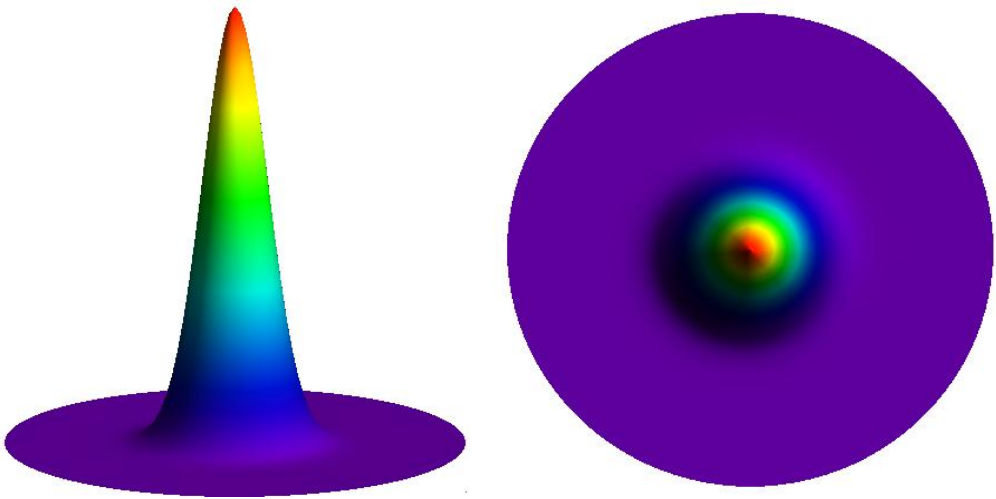

Figure 2. Graph of $W_3^{(0)}(\rho,\varphi,T/4) = W_3^{(0)}(\rho,\varphi,3T/4)$ or $W_0$

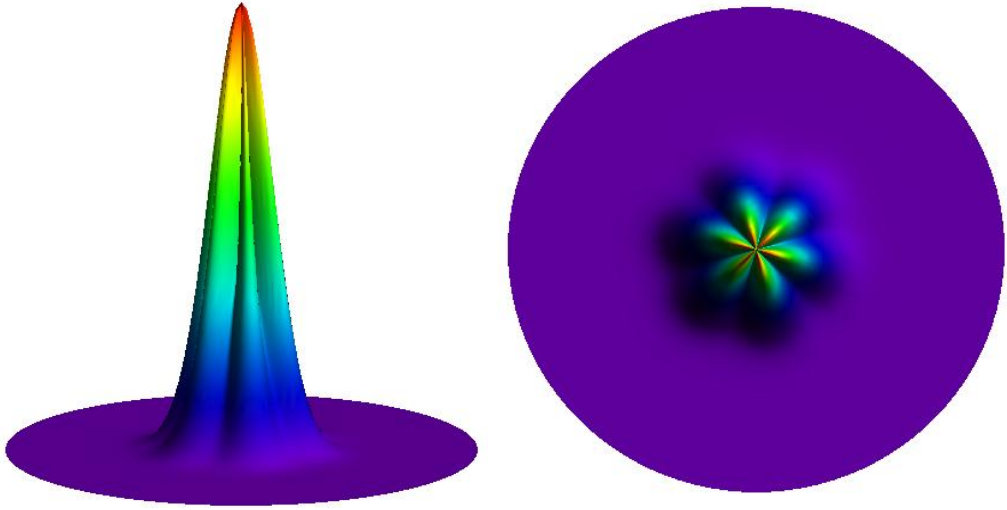

Figure 3. Graph of $W_3^{(0)}(\rho,\varphi,T/2)$

Figures 1-6 show graphs of the function (2.3) for $\ell = 3$ and $A = 2$, $C = 5$. The graphs in figures 1-3 correspond to the ground state $n = 0$, and in figures 4-6 – to the state with the number $n = 5$. In figures 1, 4 show the distributions corresponding to the point of time $t = 0$ or $t = T$. The distributions in figures 2, 5 correspond to the time $t = T/4$, when the function (2.1) becomes the ordinary Wigner function $W_n$ of the harmonic oscillator (1.9). In figures 3, 6, the distributions refer to the point of time $t = T/2$.

At the points of time $t = 0$ (figures 1,4) and $t = T/2$ (figures 3,6), the function $W_\ell^{(n)}$ deviates as much as possible from the function $W_n$ at the antinode angles $\varphi_{n,\ell}^{(anti-node)}$ (2.4). At the points of time $t = T/4$ and $t = 3T/4$, the function $W_\ell^{(n)}$ coincides completely with the function $W_n$. On the line of nodes $\varphi_{n,\ell}^{(node)}$ at all points of time, the values of the functions $W_n$ and $W_\ell^{(n)}$ coincide.

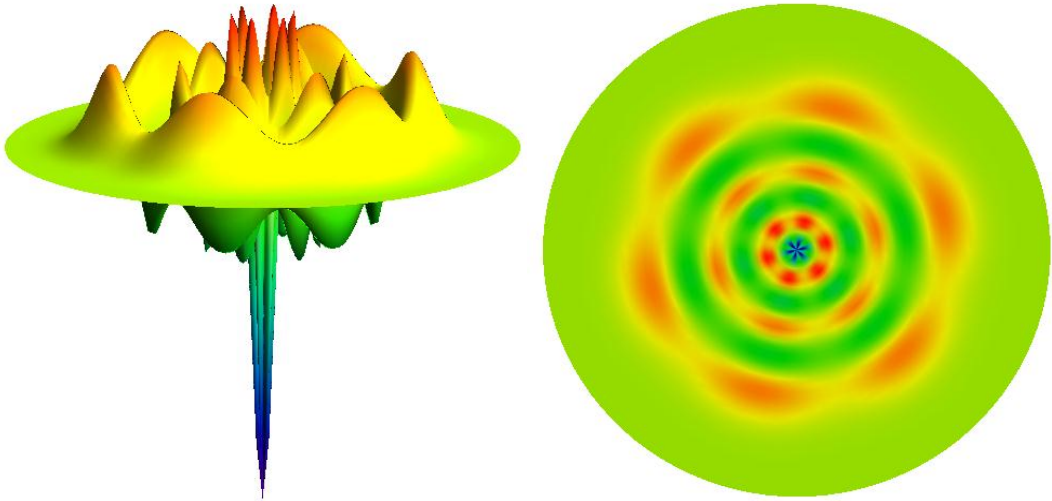

Figures 4. Graph of $W_3^{(5)}(\rho,\varphi,0) = W_3^{(5)}(\rho,\varphi,T)$

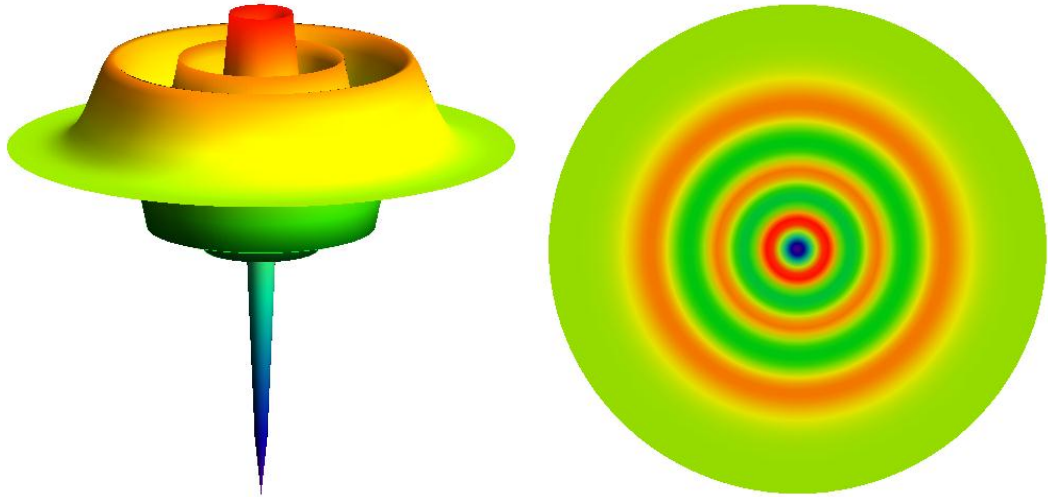

Figures 5. Graph of $W_3^{(5)}(\rho,\varphi,T/4) = W_3^{(5)}(\rho,\varphi,3T/4)$

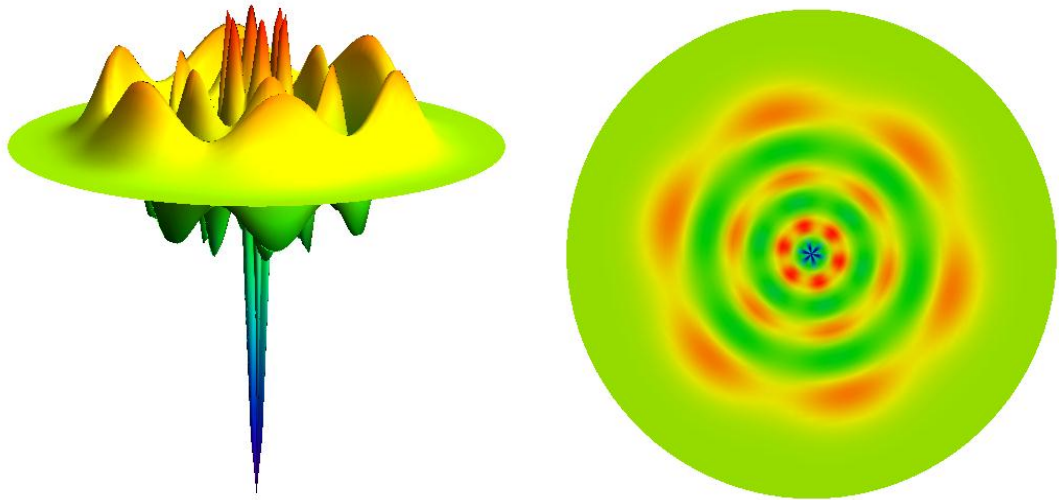

Figures 6. Graph of $W_3^{(5)}(\rho,\varphi,T/2)$

**Remark**

Note that the ground state ($n=0$) shown in figures 1-3 does not have regions of negative values of the function $W_\ell^{(0)}$, since the coefficient $2\dfrac{A}{C}$ in the expression (2.3) equals $\dfrac{4}{5}<1$. It may seem that the positivity of the function $W_\ell^{(0)}$ violates Hudson's theorem [42]. That is, there is a positive function $W_\ell^{(0)}$ different from the Gaussian distribution. In fact, there is no contradiction, since Hudson's theorem is proved for analytic functions, and the function $W_\ell^{(0)}$ is non-smooth at the origin.

**Conclusion**

Initially, the Wigner function was constructed phenomenologically in the form of a Fourier transform:

$$W(x,p,t) = \frac{1}{2\pi\hbar} \int_{-\infty}^{+\infty} \exp\left(-i\frac{ps}{\hbar}\right) \left\langle x+\frac{s}{2}\right| \hat{\rho}(t) \left|x-\frac{s}{2}\right\rangle ds, \qquad (C.1)$$

where $\hat{\rho}$ is a density matrix. Substituting the expressions for the wave functions $\Psi_n$ of the harmonic oscillator into the integral (C.1), we can obtain the functions $W_n$ (1.9). The wave functions $\Psi_n$ are stationary, therefore, the functions $W_n$ are also stationary (1.9).

Using the von Neumann equation for the density matrix $\hat{\rho}$ and the expression (C.1), the Moyal evolution equation (1.3) is obtained. Naturally, the function $W_n$ is a solution of the Moyal equation, but not any solution of the Moyal equation will give a function that has the properties (1.16). In this work, to fulfill the property (1.16), additional conditions were imposed on the function $\Phi$ (see Theorem 2).

Solving the Moyal equation in a general form is complicated by the presence of an infinite series on the right-hand side of (1.3), but this is only part of the problem. Another part of the problem is the choice of initial-boundary conditions, which, at least, must be related to the conditions (1.16). From the point of view of classical mechanics, one could consider the expression for the function $\bar{\Phi}$ in the form of a running wave:

$$\bar{\Phi}(\varphi,t) = A\cos(\Omega t - \kappa\varphi), \qquad (C.2)$$

but the «quantum» conditions (1.16) are not fulfilled for it. At the same time, solving the Schrödinger equation for the harmonic oscillator, there is no serious problem in choosing the initial-boundary conditions.

It is interesting that from the standpoint of «classical» mechanics (the Moyal, Liouville equations) we obtain the wave equation (1.5), which imposes no conditions on the dependence of the Wigner function on the radius $\rho$, i.e., on energy $\varepsilon$. Important for classical mechanics is the dependence of the Wigner function on the time $t$ and angle $\varphi$ (see the equation (1.5)). From the standpoint of quantum mechanics, the Wigner function (1.9) contains a dependence only on the energy $\varepsilon$, i.e., on the radius $\rho$, and does not depend on the angle $\varphi$.

Thus, for the same physical system (oscillator), completely «orthogonal» conditions are imposed. Quantum mechanics defines a radial (energy) dependence, and classical mechanics (Liouville, Moyal equations) imposes conditions only on time and angular dependence.

**Acknowledgements**
This work was supported by the RFBR No. 18-29-10014.

**Appendix**
*Proof of Theorem 1*

Let us introduce the following notation

$$u = \omega\bar{x}, \quad v = \frac{p}{m}, \quad W'(u,v,t) = \bar{W}(\bar{x},p,t) = W(x,p,t), \qquad (A.1)$$

then the equation (1.4) is of the form:

$$\frac{1}{\omega}\frac{\partial W'}{\partial t} + v\frac{\partial W'}{\partial u} - u\frac{\partial W'}{\partial v} = 0, \qquad (A.2)$$

Let us write the equation (A.2) in the polar coordinates:

$$u = \rho\cos\varphi, \ v = \rho\sin\varphi, \ W'(u,v,t) = W(\rho,\varphi,t), \qquad (A.3)$$

$$\frac{\partial W'}{\partial u} = W_\rho \cos\varphi - W_\varphi \frac{\sin\varphi}{\rho}, \ \frac{\partial W'}{\partial v} = W_\rho \sin\varphi + W_\varphi \frac{\cos\varphi}{\rho},$$

$$\frac{1}{\omega}\frac{\partial W}{\partial t} + \rho\sin\varphi\left(W_\rho \cos\varphi - W_\varphi \frac{\sin\varphi}{\rho}\right) - \rho\cos\varphi\left(W_\rho \sin\varphi + W_\varphi \frac{\cos\varphi}{\rho}\right) = 0,$$

$$W_t = \omega W_\varphi. \qquad (A.4)$$

The equation (A.4) corresponds to the wave equation, which is obtained from the equation (A.4) when it is differentiated with respect to time:

$$W_{tt} = \omega W_{\varphi t} = \omega^2 W_{\varphi\varphi}, \qquad (A.5)$$

which was to be proved.

## Proof of Lemma 1

Let us calculate the integral of the function (1.10) over the entire phase space:

$$dxdp = \frac{m}{\omega}dudv = \frac{m}{\omega}\rho d\rho d\varphi,$$

$$\int_{-\infty}^{+\infty}\int_{-\infty}^{+\infty} W_n(x,p)dxdp = (-1)^n \frac{2m}{\hbar\omega}\int_0^{+\infty} e^{-\frac{m}{\hbar\omega}\rho^2} L_n\left(\frac{2m}{\hbar\omega}\rho^2\right)\rho d\rho = 1,$$

$$\frac{1}{N} = \frac{(-1)^n m}{\pi\hbar\omega}\int_0^{2\pi}\left[C + f(\Omega t + \kappa\varphi) + g(\Omega t - \kappa\varphi)\right]d\varphi \int_0^{+\infty} e^{-\frac{m}{\hbar\omega}\rho^2} L_n\left(\frac{2m}{\hbar\omega}\rho^2\right)\rho d\rho,$$

$$\frac{1}{N} = C + \frac{1}{2\pi}\int_0^{2\pi}\left[f(\Omega t + \kappa\varphi) + g(\Omega t - \kappa\varphi)\right]d\varphi = C + \langle f \rangle + \langle g \rangle,$$

which was to be proved.

## Proof of Theorem 2

Since the function $W_n(\bar{x},p)$ is even and the unction $\Phi(\bar{x},p,t)$ is odd in respect to the variables $\bar{x}$ and $p$, the average values of $\langle \Phi \rangle$ (1.15) will be equal to zero, i.e.

$$\langle \Phi \rangle_{p,n}(\bar{x}_s,t) = \langle \Phi \rangle_{x,n}(p_s,t) = 0. \qquad (A.6)$$

Let us calculate the value of the normalization factor $N$, which, in accordance with the expression (1.11), depends on the mean values of the functions $\langle f \rangle$ and $\langle g \rangle$. The integration over the momentum $p$ from $-\infty$ to $+\infty$ corresponds to the integration over the angle $\varphi \in \left[-\dfrac{\pi}{2}, \dfrac{\pi}{2}\right]$. The angle $\varphi$ can be presented in the form:

$$\varphi = \begin{cases} \operatorname{arctg} \dfrac{p}{\bar{x}}, & \bar{x} \geq 0, \\ \pi + \operatorname{arctg} \dfrac{p}{\bar{x}}, & \bar{x} < 0. \end{cases} \quad (A.7)$$

Taking into account (A.6) and (A.7), we obtain:

$$d\varphi = \dfrac{\bar{x} dp}{\bar{x}^2 + p^2}, \quad dp = \dfrac{\rho}{\cos \varphi} d\varphi = \bar{x} \dfrac{d\varphi}{\cos^2 \varphi},$$

$$0 = \bar{x} \int_{-\pi/2}^{\pi/2} \Phi(\bar{x}, \bar{x} \operatorname{tg} \varphi, t) \dfrac{W_n(\bar{x}, \bar{x} \operatorname{tg} \varphi)}{\cos^2 \varphi} d\varphi, \quad (A.8)$$

where the value $\bar{x}$ is fixed when integrating over the angle $\varphi$. On the interval $\varphi \in \left[-\dfrac{\pi}{2}, \dfrac{\pi}{2}\right]$, the function $\dfrac{W_n(\bar{x}, \bar{x} \operatorname{tg} \varphi)}{\cos^2 \varphi}$ is an even function with respect to the point $\varphi = 0$ and $\varphi = \pi$ ($p = 0$), therefore the equality of the integral (A.8) to zero is caused by the oddness of the function $\Phi(\bar{x}, \bar{x} \operatorname{tg} \varphi, t)$, i.e.

$$\int_{-\pi/2}^{\pi/2} \Phi(\bar{x}, \bar{x} \operatorname{tg} \varphi, t) d\varphi = \int_{-\pi/2}^{\pi/2} \left[ f(\Omega t - \kappa \varphi) + g(\Omega t + \kappa \varphi) \right] d\varphi = 0. \quad (A.9)$$

Let us make a substitution of the variables $\chi = \varphi + \pi$ in the integral (A.9), we obtain

$$\int_{\pi/2}^{3\pi/2} \left[ f(\Omega t_1 - \kappa \chi) + g(\Omega t_2 + \kappa \chi) \right] d\chi = 0, \quad (A.9)$$

$$\Omega t_1 = \Omega t + \kappa \pi, \quad \Omega t_2 = \Omega t - \kappa \pi,$$

The phase difference between the angles $\Omega t_1$ and $\Omega t_2$ is $2\pi\kappa$, i.e. $\Omega t_1 = \Omega t_2 + 2\pi\kappa$. The functions $f$ and $g$, by virtue of (1.7), are $2\pi\kappa$ periodic, therefore, from the expression (A.9) we obtain

$$\int_{\pi/2}^{3\pi/2} \left[ f(\Omega t - \kappa \varphi) + g(\Omega t + \kappa \varphi) \right] d\varphi = 0. \quad (A.10)$$

Adding the expressions (A.9) and (A.10), we obtain

$$\dfrac{1}{2\pi} \int_0^{2\pi} \left[ f(\Omega t - \kappa \varphi) + g(\Omega t + \kappa \varphi) \right] d\varphi = \langle f \rangle + \langle g \rangle = 0,$$

$$N = 1/C.$$

The theorem is proved.

*Proof of Theorem 3*

Let us find the energy levels $\langle\langle \varepsilon_n \rangle\rangle$ for the distributions $W^{(n)}$ (1.10):

$$d\rho^2 = 2\rho d\rho = 2\frac{\hbar\omega}{m}d\varepsilon, \quad \rho d\rho = \frac{\hbar\omega}{m}d\varepsilon, \quad d\bar{x}dp = \frac{m}{\omega}\frac{\hbar\omega}{m}d\varepsilon d\varphi = \hbar d\varepsilon d\varphi,$$

$$\langle\langle \varepsilon_n \rangle\rangle = \hbar\omega \int_{-\infty}^{+\infty}\int_{-\infty}^{+\infty} \varepsilon(\bar{x},p) W^{(n)}(\bar{x},p) d\bar{x}dp =$$

$$= N\hbar^2\omega \int_0^{2\pi}\left[C + f(\Omega t + \kappa\varphi) + g(\Omega t - \kappa\varphi)\right]d\varphi \frac{(-1)^n}{\pi\hbar} \int_0^{+\infty} e^{-2\varepsilon} L_n(4\varepsilon)\varepsilon d\varepsilon =$$

$$= N\hbar^2\omega \frac{2n+1}{4\pi\hbar} \int_0^{2\pi}\left[C + f(\Omega t + \kappa\varphi) + g(\Omega t - \kappa\varphi)\right]d\varphi,$$

$$\langle\langle \varepsilon_n \rangle\rangle = \hbar\omega\left(n+\frac{1}{2}\right)\frac{N}{2\pi}\int_0^{2\pi}\left[C + f(\Omega t + \kappa\varphi) + g(\Omega t - \kappa\varphi)\right]d\varphi, \qquad (A.6)$$

where [45] is taken into account $\int_0^{+\infty} e^{-2\varepsilon} L_n(4\varepsilon)\varepsilon d\varepsilon = (-1)^n \frac{2n+1}{4}$. Given the normalization condition (1.11) (Lemma 1), the expression (A.6) becomes:

$$\langle\langle \varepsilon_n \rangle\rangle = \hbar\omega\left(n+\frac{1}{2}\right),$$

which was to be proved.